\documentclass{vgtc}                          

\ifpdf
  \pdfoutput=1\relax                   
  \usepackage{graphicx}                
  \DeclareGraphicsExtensions{.pdf,.png,.jpg,.jpeg} 
\else
  \usepackage{graphicx}                
  \DeclareGraphicsExtensions{.eps}     
\fi%

\graphicspath{{figures/}{pictures/}{images/}{./}} 

\newcommand{\ak}[1]{\textcolor{black}{#1}}

\usepackage{microtype}                 
\PassOptionsToPackage{warn}{textcomp}  
\usepackage{textcomp}                  
\usepackage{mathptmx}                  
\usepackage{times}                     
\usepackage{cite}                      
\usepackage{tabu}                      
\usepackage{booktabs}                  
\usepackage{wrapfig}
\usepackage{tabularx}
\usepackage{enumitem}
\usepackage{subfigure}
\usepackage{xspace}
\newcommand{\etal}{\emph{et al.}\@\xspace}
\newcommand{\ie}{\emph{i.e.}\xspace}
\newcommand{\eg}{\emph{e.g.}\xspace}

\usepackage{enumitem}

\setlist[enumerate]{itemsep=0.5ex,parsep=0pt}

\onlineid{1140}

\vgtccategory{Research}

\vgtcinsertpkg

\title{Taken By Surprise? Evaluating how \ak{Bayesian} Surprise \& Suppression Influences Peoples’ Takeaways in Map Visualizations}

\author{Akim Ndlovu\thanks{e-mail: andlovu@wpi.edu} %
\and Hilson Shrestha\thanks{e-mail: hshrestha@wpi.edu} %
\and Lane T. Harrison\thanks{e-mail: ltharrison@wpi.edu}}
\affiliation{\scriptsize \ak{Worcester Polytechnic Institute}}

\teaser{
  \centering
  \includegraphics[width=\linewidth]{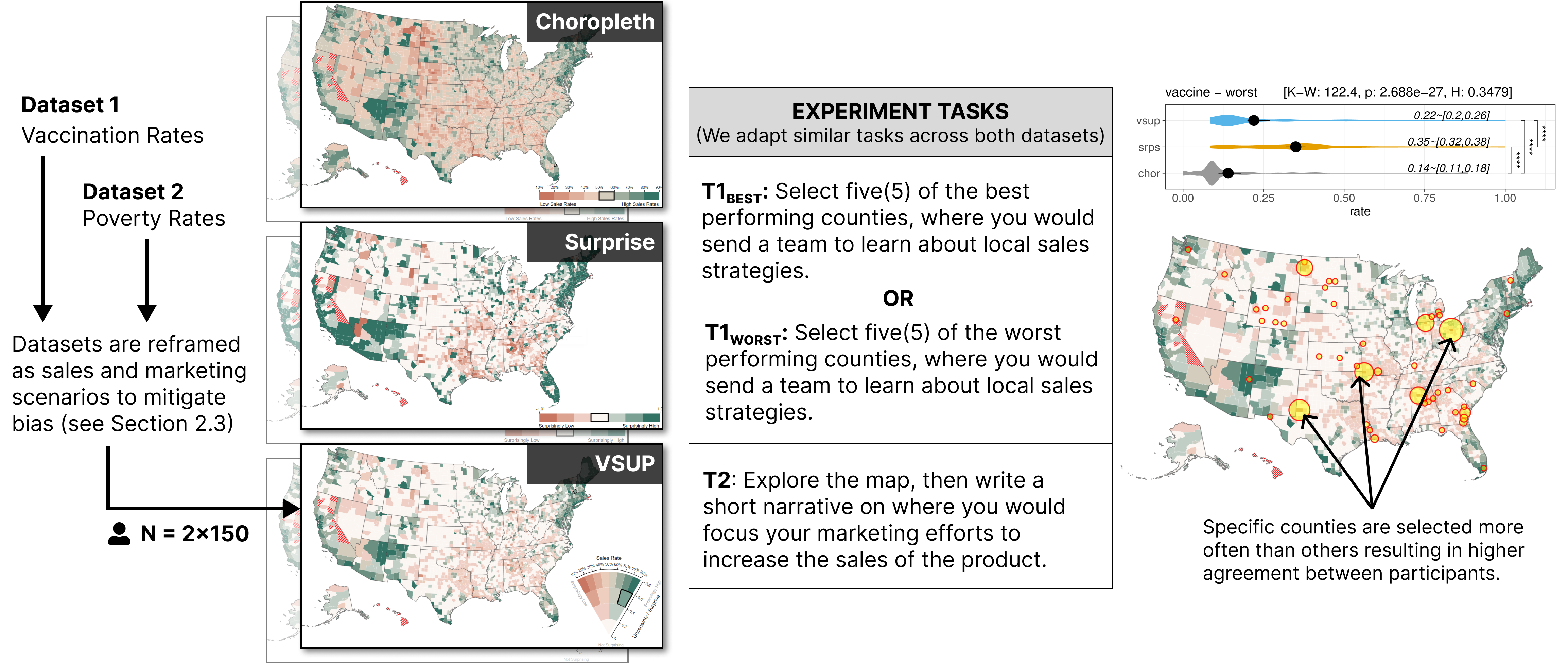}
 
  \ak{\caption{How do Bayesian surprise metrics and suppression encodings influence peoples' takeaways in map visualizations? We conduct two experiments with Covid-19 and Poverty datasets, randomly assigning 300 participants to three map conditions. We collect data across three map takeaway tasks \textit{T1\textsubscript{Best}, T1\textsubscript{Worst}: Identify} and \textit{T2: Explore}. To mitigate biases for particular dataset contexts discovered in pilot studies (\eg vaccine skepticism) we reframe both datasets as a sales and marketing task. Metrics include participants' exploration metadata (quantitative) and takeaway comments (qualitative).}}
  \label{fig:teaser}
}

\abstract{
Choropleth maps have been studied and extended in many ways to counteract the many biases that can occur when using them. 
Two recent techniques, Surprise metrics and Value Suppressing Uncertainty Palettes (VSUPs), offer promising solutions but have yet to be tested empirically with users of visualizations. 
In this paper, we explore how well people can make use of these techniques in map exploration tasks. 
We report a crowdsourced experiment where $n=300$ participants are assigned to one of Choropleth, Surprise (only), and VSUP conditions (depicting rates and Surprise in a suppressed palette). 
Results show clear differences in map analysis outcomes, \eg with Surprise maps leading people to significantly higher areas of population, or VSUPs performing similar or better than Choropleths for rate selection. 
Qualitative analysis suggests that many participants may only consider a subset of the metrics presented to them during exploration and decision-making. 
We discuss how these results generally support the use of Surprise and VSUP techniques in practice, and opportunities for further technique development.
\ak{The material for the study (data, study results and code) is publicly available on \url{https://osf.io/exb95/}}.
} 

\CCScatlist{
  \CCScatTwelve{Data Visualization}{Choropleth Maps}{Baye\-si\-an Surprise}{}
}

\begin{document}


\firstsection{Introduction}
\maketitle
The vast amount of data gathered during Covid-19 pandemic has created a need to visualize and accurately communicate trends in vaccinations, deaths and infections \cite{griffin2020trustworthy,juergens2020trustworthy}. As a result, Choropleth maps have been widely used for visualizing trends in geospatial data such as high or low performing regions and regions that show a high degree of correlation or disparity \cite{munn2020developing, mooney2020mapping}.
However, research has shown that when visualizing data that closely resembles a population distribution, Choropleth maps are prone to biases. Such instances occur  when visualizing percentage rates, where counties or regions with low population and high variance may be shaded using darker colors, which may be misleading for map readers \cite{correll2016surprise}.

A number of approaches have been proposed to counteract bias in Choropleth maps, resulting in a modified or supplemented dataset~\cite{correll2016surprise}. These include normalization, Bayesian \ak{surprise}, Spatial smoothing and Geographical weighted regression \cite{brunsdon1998geographically, genebes2018spatial,correll2016surprise}. Prior studies by MacEachren \cite{maceachren1992visualizing} have interrogated the impact of using different metrics to offset biases in map visualizations. \ak{For example, Correll and Heer use Bayesian surprise \cite{correll2016surprise} to depict a metric that measures belief about the observed data, either instead of or alongside the actual data items themselves. Studies suggest that the visualization of uncertainty requires people to understand the metrics to effectively use them \cite{hullman2019survey}, implying a need to investigate how people interpret metrics such as Bayesian surprise in map reading contexts.}

As a result, researchers have examined some model-driven mapping techniques by designing information retrieval, comparison, ranking and aggregation tasks, in order to understand their impact on pattern recognition and decision making \cite{maceachren1998visualizing, blenkinsop2000evaluating, boukhelifa2012evaluating, alberti2018web}. Other approaches for evaluating such maps include using empirically derived frameworks similar to the one proposed by Roth~\cite{roth2013empirically}. 
Although widely applicable to map evaluation studies, such frameworks may need to be extended when uncertainty is added as a consideration \cite{hullman2019survey}.

Two recently developed techniques provide a promising baseline for investigating map debiasing techniques in user studies.
Correll and Heer propose the use of ``Surprise'', a Bayesian weighting technique that offsets biases in map visualizations \cite{correll2016surprise}. 
Surprise up-weights or down-weights data points that deviate from expected values, by calculating an updated belief about the data based on prior knowledge. 
Correll \etal \cite{correll2018value} also introduce the Value-Suppressing Uncertainty Palette, a map coloring technique and legend technique which can visualize both uncertainty measures (such as Surprise) and rates in a single map.

\ak{In this paper, we examine the impact of two recently developed visualization techniques, on how people explore and generate takeaways in map reading contexts. 
We report a crowdsourced study with $n=300$ participants}, where we ask participants to perfom map analysis tasks with one of three visualization conditions: Choropleth, Surprise, or Value Suppressing Uncertainty Palettes (VSUPs). 
We describe some of the technical challenges and resulting adaptations in taking prior map tasks and task taxonomies to study techniques which emphasize different metrics (Surprise, rates, or both).
In particular, we leverage previous research by Roth \cite{roth2013empirically} and Besan\c{c}on et~al. \cite{besanccon2020evaluation} to design a universal task for all map conditions (see \autoref{tab:taskList}). 
Results show clear differences in map analysis outcomes (\autoref{fig:quantative-analysis}),
while qualitative analysis suggests that participants in some cases only consider a subset of the metrics available to them. 
We discuss how these results tentatively support the use of Surprise and VSUP techniques for broader visualization viewing populations, while also highlighting challenges that might be addressed through future design and technique development.

\section{METHODOLOGY}
We designed three interactive stimuli (Choropleth, Surprise, and VSUP maps) using Covid-19 Vaccination and Poverty datasets. 
We conducted two experiments on the online crowdsourcing platform Prolific, where we collected data from $n = 300$ participants. 
\ak{Pilot studies using a vaccine dataset revealed skewed results with strong political bias (\S{} \ref{pilot_study})}. 
We therefore design a scenario that ``masks'' the underlying dataset as being about sales rates, using tasks adapted from Roth \cite{roth2013empirically} and Besan\c{c}on \cite{besanccon2020evaluation}. 
\ak{To test for the possible impact of data characteristics, we repeat the experiment across two datasets measuring different geospatial phenomena, Covid-19 vaccination rates and poverty statistics}.

\subsection{Stimuli Design}
Our design goal was to minimize notable differences between the stimuli to avoid map interpretation bias, while maximizing on techniques that improve the accessibility of information \cite{latif2021deeper, robinson2017geospatial, andrienko1999interactive}. 
For ecological validity, our map design and color schemes were influenced by The New York Times (NYT) Covid-19 vaccination map \cite{Aisch2015-pl}, and designed to be consistent as possible between all three maps \ak{(see supplemental material for additional stimuli design considerations).}

\subsection{Experiment Datasets}
We adapted publicly available county level datasets of Covid-19 vaccinations \cite{cdc} and Poverty rates \cite{openintro} of the US. 
Prior to conducting the study, we replicated a Surprise map of per-capita unemployment rates from Correll and Heer \cite{correll2016surprise}, that uses a model of the deMoivre's funnel to determine deviations from the average per-capita rate.
This method calculates the test statistic ($Z_{s}$) from event rates. 
Bayesian methods are then used to find the likelihood of points being $Z_{s}$ distant from the center of the funnel: 

\begin{equation}
{\large P(s|deMoivre) = 1 - (2 \cdot \int_{0}^{|Z_{s}|} \phi(x) dx)}
\end{equation}
where deMoivre represents the model and {$s \in D$} (Dataset).
After replicating the Surprise map of per-capita unemployment rate, we apply the same process to our datasets of interest \cite{cdc, openintro}.

\begin{figure*}
   \includegraphics[width=\textwidth]{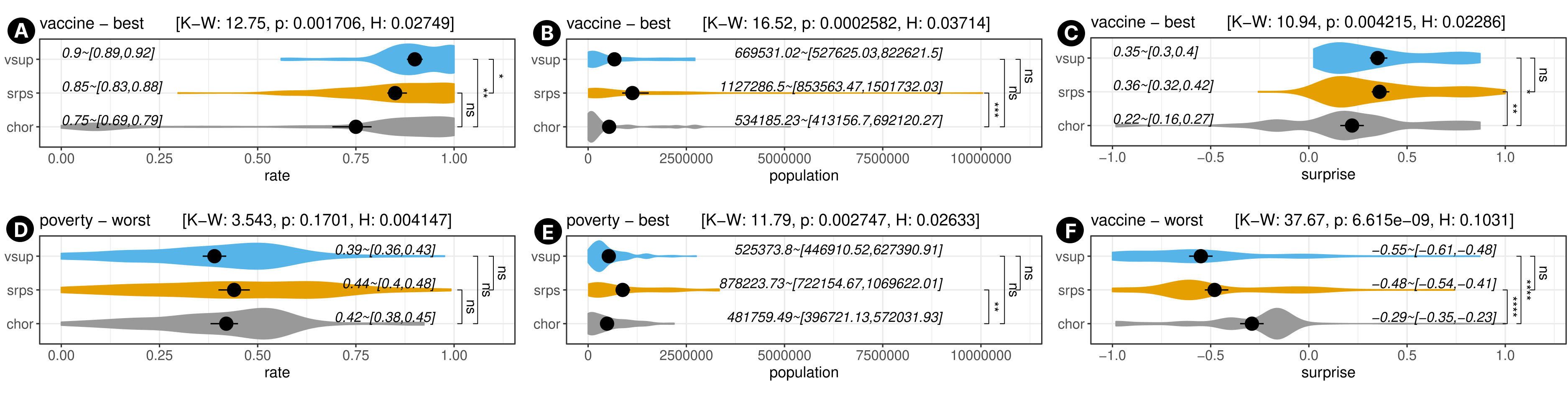}
   \caption{Quantitative results for rate, population and surprise metrics. \textbf{A)} T1\textsubscript{Best}-Vaccine (Rate) \textbf{B)} T1\textsubscript{Best} (Population) \textbf{C)} T1\textsubscript{Best}-Vaccine (Surprise) \textbf{D)} T1\textsubscript{Worst}-Poverty (Rate) \textbf{E)} T1\textsubscript{Best}-Poverty (Population) \textbf{F)} T1\textsubscript{Worst}-Vaccine (Surprise). We report Kruskal-Wallis tests on overall effects and post-hoc tests (brackets). $95\%$ confidence interval using a bootstrap method are depicted. Our findings suggest that VSUPs can lead to the selection of counties with higher rates and surprise, and Surprise maps lead to the selection of highly populated counties.}
   \label{fig:quantative-analysis}
\end{figure*}

\subsection{Pilot Study}
\label{pilot_study}
To refine the user experience for the study, we conducted a pilot study with $n=30$ participants. 
We designed our stimuli using a Covid-19 dataset \cite{cdc} and randomly assigned $n = 10$ participants to each condition. 
Our initial analysis of the qualitative feedback reflected a high degree of participants' personal beliefs and political affiliation. Here is one example: 

\textbf{Response:} \textit{``I think it's going okay. In the beginning everyone was reluctant since it's so new, there's hardly and research. [...] It seems like there'll be a lot more people vaccinated by the end of the year"}.

Given these results, we rephrased our tasks to a product sales and marketing decision-making problem using both the Covid-19 and Poverty datasets (see \autoref{tab:taskList}). 
\ak{We developed two task categories across the metrics and conditions to be considered by summarizing map analyses objectives and tasks used in prior studies from Roth \cite{roth2013empirically} and Besan\c{c}on \cite{besanccon2020evaluation}}.
We also developed additional ``scrollytelling'' trainings for all conditions to help reduce sources of noise (\eg participant misunderstandings) during the full experiment.

\subsection{Task and Procedure}
We used a between subjects design across two experiments (Covid-19 and Poverty). 
We designed 3 stimuli (conditions) and 3 tasks $T1_{Best}$, $T1_{Worst}$ and $T2_{Explore}$ (see \autoref{fig:teaser}$1$). 
We randomly assigned $25$ participants to each condition. 
The total number of participants for the study was therefore, $2$ experiments $\times 3$ conditions (Choropleth, VSUP and Surprise stimuli) $\times 2$ tasks ($T1_{Best}$, $T1_{Worst}$) $\times 25$ participants $= 300$. 
\ak{Of our participants, 160 identified as female, 136 identified as male, and 4 participants chose not to disclose their gender. 
Participants' age ranged from 18 to 76 with an average of 35.}
The study was IRB-reviewed and we required a consent form before participation.
Participants were not constrained to a completion time, however, we estimated an average completion time of 7 minutes, used to calculate a payment of \$1.40 to exceed US Minimum Wage. 
We collect metadata on counties of interest for each participant (\eg population), as well as feedback regarding their perception of the study.

\begin{table}
    \caption{List of experiment tasks \textit{T1: Identify} and \textit{T2: Explore}}
    \label{tab:taskList}
    \begin{tabularx}{\linewidth}{>{\hsize=0.1\hsize}X>{\hsize=0.20\hsize}X>{\hsize=0.70\hsize}X}
        \hline
        & Objective & Task Narration \\
        \hline
        T1\textsubscript{Best} & Identify \newline and Rank & Select five (5) of the best performing counties, where you would send a team to learn about local sales strategies. \\
        \hline
        T1\textsubscript{Worst} & Identify \newline and Rank & Select five (5) of the worst performing counties, where you would send a team to learn about local sales strategies. \\
        \hline
        T2 & Compare \& \newline Delineate \newline (Explore) & Explore the map, then write a short narrative on where you would focus your marketing efforts to increase sales of the product. \\
        \hline
    \end{tabularx} 
\end{table}

\section{Results}

We used a Kruskal-Wallis test to detect overall effects in data across the three different mapping techniques (see \autoref{fig:quantative-analysis}). 
For post hoc tests, we use Dunn's test with Bonferroni correction. 
We also compute and report $95\%$ confidence intervals using bootstrapping.
For geospatial analysis, we create a point map of participant county selections across the three conditions (see \autoref{fig:pselections}).

\subsection{Identify Tasks \texorpdfstring{T1\textsubscript{Best} and T1\textsubscript{Worst}}{T1Best and T1Worst}}

\ak{In both experiments (Covid-19 and Poverty datasets), participants' performance differed across the tested visualization conditions in terms of rate, population, and Surprise metrics of the selected counties (see \autoref{fig:quantative-analysis} and supplemental material for full results)}.

\textbf{Rate:}
\ak{We find overall differences between the map conditions for vaccine best $KW=12.75$ $p=0.0017$ $H=0.02749$ (see \autoref{fig:quantative-analysis}A) and the vaccine worst tasks $KW=122.4$ $p=2.688e-27$ $H=0.3479$.
Post-hoc comparisons suggest that the VSUP performs best in the vaccine best task, and the Choropleth map performs best in the vaccine worst task. 
In the latter case, VSUPs appear to balance the differences between the Surprise and Choropleth maps, making them a potentially good choice overall.
However, we note that all maps performed similarly when selecting the worst performing counties in the poverty dataset \autoref{fig:quantative-analysis}D).
While a formal method for investigating dataset distributions would be needed, it appears that the poverty rates in the dataset itself are negatively skewed, which may be a reason for the observed similar performance, implying a need for more distribution-sensitive techniques in future work.}

\textbf{Population:}
In terms of selected counties, the Surprise maps tended to lead participants towards counties of higher population (see \autoref{fig:quantative-analysis}B and \autoref{fig:quantative-analysis}E).
In particular, we find in the vaccine best task an overall effect $KW=16.52$ $p=0.00025$ $H=0.037$, and for in the poverty best task $KW=11.79$ $p=0.00275$ $H=0.026$.
However, we note that these effects tend to place Surprise maps above Choropleth maps, but not above VSUPs, which appear to balance the effects of the other two.
Similar effects and trends are found in the \ak{vaccine worst $KW=107.4$ $p=4.695e-24$  $H=0.3047$, and in the poverty worst $KW=16.3$ $p=0.0002798$ $H=0.03861$ tasks}.

\textbf{Surprise:} 
Results suggest that VSUPs and Surprise maps led participants to select counties with high surprise values for the vaccine best and low surprise values for the vaccine worst task as shown in \autoref{fig:quantative-analysis}C with $KW=10.94$ $p=0.004$ $H=0.02286$ and \autoref{fig:quantative-analysis}F with $KW=37.67$ $p=6.615e-09$ $H=0.1031$. 
However, in the poverty best and worst tasks, while overall effects were observed $KW=14.23$ $p=0.0008$ $H=-0.033$ and $KW=51 $ $p=8.01e-12$ $H=0.1319$, these generally indicate VSUPs outperforming Surprise and Choropleth maps. 
These differences again may be partially due to the skewed nature of the poverty rates compared to the vaccine rates.

\subsection{Explore Task T2}
\begin{figure*}[t]
  \includegraphics[width=\textwidth]{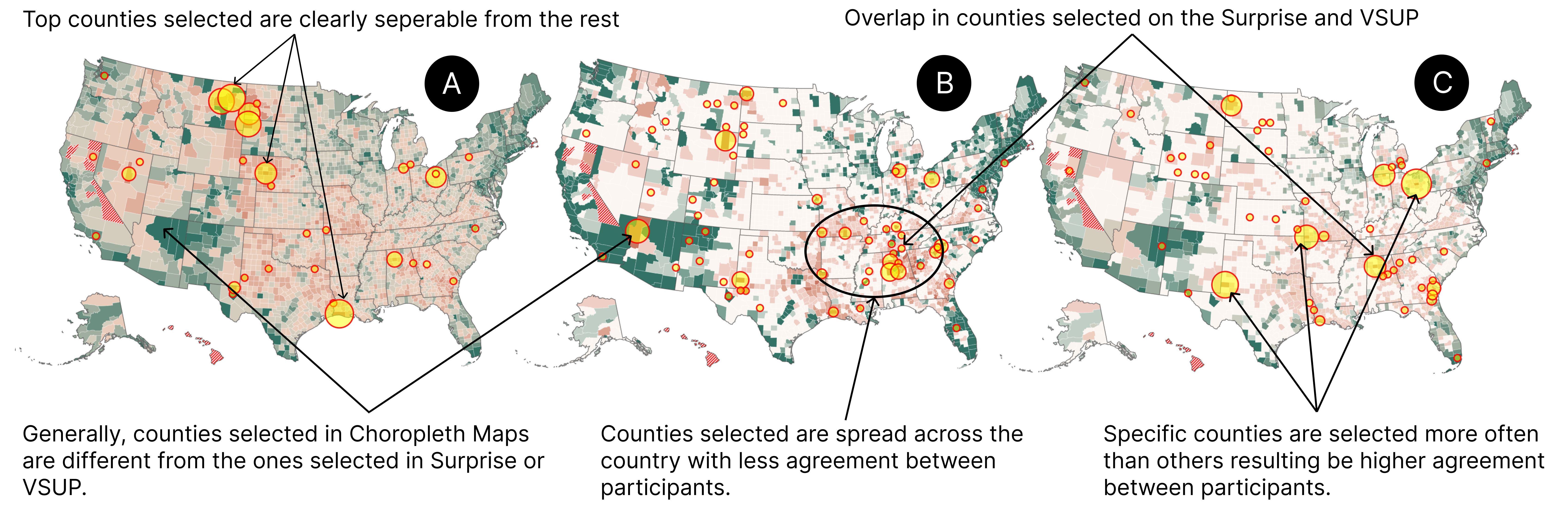}
\caption{Visual Analysis of participants' county selection for condition T1\textsubscript{Worst}. \textbf{A)} Choropleth \textbf{B)} Surprise \textbf{C)} VSUP maps. We conducted a visual analysis of participants' selections. We find consistent county selections on the VSUP, whereas Surprise map has greater dispersion. However, we see a higher degree of correlation in participants' selections between the VSUP and Surprise maps, compared to the Choropleth map.}
  \label{fig:pselections}
\end{figure*}

We considered participants' feedback based on relevance, similarity and identified keywords such as population, color, surprisingly high and surprisingly low. 
In the Discussion we expand on our takeaways from participant responses which suggest that: 
\begin{enumerate}
  \item Participants only consider a subset of the metrics presented (\S{} \ref{subsubdescription_takeaway}).
  \item \ak{Visual encodings (\eg color) can impact how people interpret surprise} (\S{} \ref{subsubcolor_surprise}).
  \item County size can skew peoples takeaways (\S{} \ref{subsubsequential_uncertainty}).
\end{enumerate}

\section{DISCUSSION}
\subsection{Spatial Analysis of Identify Task \texorpdfstring{T1\textsubscript{Best} and T1\textsubscript{Worst}}{T1Best and T1Worst}}

Results from participant ranking selections are aggregated by county and shown in \autoref{fig:pselections}. 
We infer the following takeaways from participants' interactions and selections on the maps.

\subsubsection{Visual Saliency of High/Low performing counties} 
County selection on the Choropleth map show a high level of dissimilarity compared to the Surprise and VSUP maps. 
We attribute this to the narrowed visual search space when visualizing Surprise compared to event-rates on a Choropleth map. 
This is supported by a further analysis of the Surprise and VSUP maps, where we see tighter clusters and consensus of participants ranking selections. 
Results also show a higher degree of ranking consensus by participants' on the VSUP map compared to the Surprise map. 
This may be due to the fact that VSUPs further suppress highly uncertain values by combining color cells in a palette using a tree structure \cite{kay2019much}. 
To assess whether participants considered population in their decision-making process, we conducted further analysis of the population of counties they selected. 
Population focused results suggest that participants consider highly populated areas when making task based decisions using both Surprise and VSUP maps. 
Such patterns are also validated by the qualitative analysis (T2).

\subsection{Explore Task T2}
We summarize feedback from an open-ended response task, where we ask participants to explore the map and give insights on where they would focus efforts to ``increase sales of the product''.

\subsubsection{Participants only consider a subset of the metrics presented on the maps.}
\label{subsubdescription_takeaway}
Participant comments suggest that some appeared to have difficulty in making use of all the available metrics (Surprise, Rate and Population), instead they used only 1 or 2 of the available metrics to select high or low performing counties. 
These findings are supported by summarizing participants' feedback on the strategies they used in selecting counties on the maps and contribute to insights on the challenges associated with comprehending Surprise without the consideration of other metrics (Population and Rate), for example:  

\textbf{Response: }\textit{``I looked for areas with high Surprise metrics (or low) and considered that most areas could be converted because of their proximity to areas with good sales"}

However, other participants effectively used interactions to explore smaller counties by hovering over the legend and counties. 
This allowed them to carry out more complex queries on the maps, suggesting that they could gain more insights by carrying out other tasks, for example:

\textbf{Response: }\textit{``[...] I'd hover over these areas to understand the surprise metric relative to the sales success rate and population. Being able to compare the data helped me to understand what the surprise metric meant, and then helped me develop a hypothesis on why these are high success/high surprise areas."}

\subsubsection{Color influences how people interpret surprise}
\label{subsubcolor_surprise}
We observed the influence of color in how participants interpret either event-rates or surprise. 
These findings suggest that some participants consider dark green and dark brown as high or low sales rate counties respectively \cite{schiewe2019empirical}. 
For example: 

\textbf{Response:}\textit{``Darker green colors show 
positive and more response to the marketing and the darker pink color is the opposite [...]"}

While interpretation is true for standard Choropleth maps, it is not  necessarily true for Surprise and VSUP maps, which depict more complex metrics. 
This may indicate a need for additional training methods or investigation of visual cues that help people associate depicted colors with metrics rather than rates alone.

\subsubsection{Size influences how people interpret of uncertainty}
\label{subsubsequential_uncertainty}
Similar to findings by Schiewe \cite{schiewe2019empirical}, both qualitative and point pattern analysis suggest that some participants neglect smaller counties and are drawn to larger counties or states on the maps. 
However, our findings also suggest that VSUPs and Surprise maps suppress low-population counties with high rates (\eg unsurprising), which may help alleviate one aspect of this bias.
How to ensure small yet high population counties also receive sufficient attention remains a challenge for maps geared towards the general public.

\section{LIMITATIONS and FUTURE WORK}
Our analysis of the Surprise, Rate and Population metrics shows clear differences between the mapping techniques used in this study (Choropleth, Surprise and VSUP maps). 
However, we hypothesize that the use of datasets with different distributions (normal and left skewed) may impact findings of our study. 
Future experiments may investigate directly the impact of skews in rates (\eg through simulation) on participant exploration and takeaways.
Furthermore, metrics and interaction techniques that build on existing work like Surprise and VSUPs may further enrich map analysis for the public.

Another limitation is noise in the experiment. 
While the sales scenario worked well overall by allowing us to ask the same task across Choropleth, Surprise, and VSUP
maps, one key issue arose in the ``worst'' tasks. We observed outlier participants across all conditions who, when prompted to select the worst performing counties, instead selected the best performing counties. 
This may be a bias with the framing of sales, which could be addressed by experimenting with other scenarios or by additional design or feedback mechanisms. 

Future work should consider collecting prior probability distributions from participants \cite{Aisch2015-pl, kim2021bayesian}. 
Experimenting with other representational techniques such as map pairs could also assist in improving the accessibility of highly technical thematic maps for the general public. 
VSUPs use a heuristic approach to suppress values at high level of uncertainty, therefore, future research could also consider the use of decision based models as suggested in work by Kay \cite{kay2019much} and Yang \etal \cite{yangsubjective}.  

\section{CONCLUSION}
Despite the pervasive use of choropleth map visualizations, especially when communicating critical data to the public (\eg vaccine trends or election results), they suffer from well-documented biases and limitations.
In this study, we design an experiment to test two recently proposed techniques, Surprise maps and VSUPs, in a crowdsourced setting similar to how participants might encounter such maps online.
Results generally indicated that Surprise maps and VSUPs do indeed offset some of the issues of traditional Choropleth maps. 
However, close inspection also reveals opportunities for addressing confusion and misconceptions of these new techniques.
Going forward, designers may benefit from knowing that Choropleths perform similarly to these new techniques (\ie reducing the risk of harm), while results that indicate  these new techniques can lead people to more surprising or populous counties may give designers the confidence to experiment with new and innovative ways of communicating with the general public.

\newpage
\bibliographystyle{abbrv-doi}

\bibliography{manuscript}
\end{document}


\onecolumn{}
\maketitle

\section{Sup-1: Participants' ranking selections (Vaccination Dataset)}
\begin{figure}[h]
    \centering
    \includegraphics[width=\textwidth]{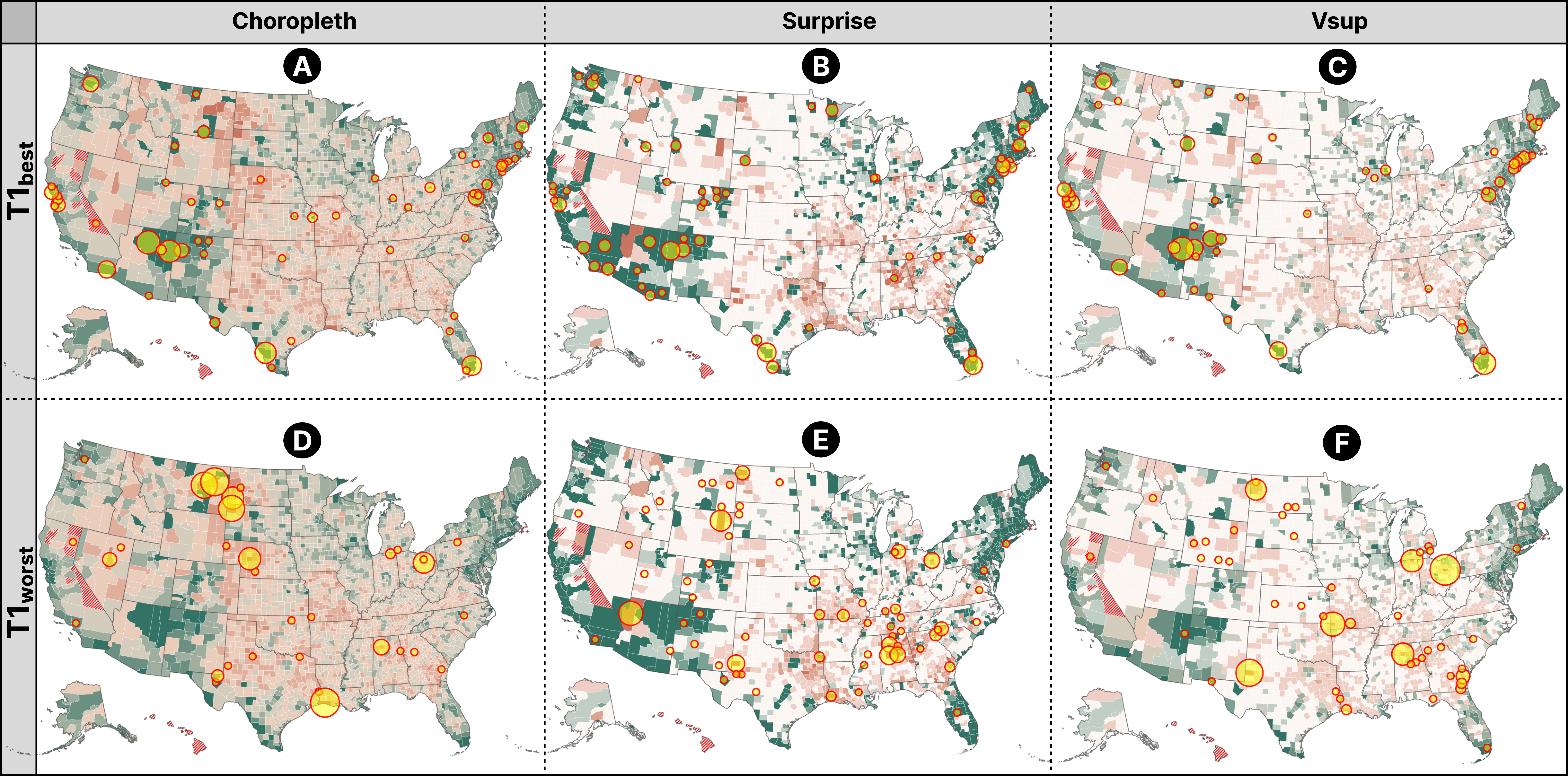}
    \caption{Participants' county selections for vaccination data tasks T1\textsubscript{Best} and T1\textsubscript{Worst}. \textbf{A)} Choropleth map (T1\textsubscript{Best}) \textbf{B)} Surprise map (T1\textsubscript{Best}) \textbf{C)} VSUP (T1\textsubscript{Best}) \textbf{D)} Choropleth map (T1\textsubscript{Worst}) \textbf{E)} (T1\textsubscript{Worst}) Surprise map \textbf{F)} VSUP (T1\textsubscript{Worst}). Visual analysis shows a high degree of consensus on the VSUP maps, particularly in \textbf{F} (VSUP). We see some consensus on \textbf{D} (Choropleth-Worst) compared to \textbf{A} (Choropleth-Best). We also see a high degree of dispersion on the Surprise maps \textbf{B} and \textbf{E} compared to both the Choropleth and VSUPs.}
    \label{fig:supplement-map}
\end{figure}

\newpage

\section{Sup-2: Participants' ranking selections (Poverty Dataset)}
\begin{figure}[h]
    \centering
    \includegraphics[width=\textwidth]{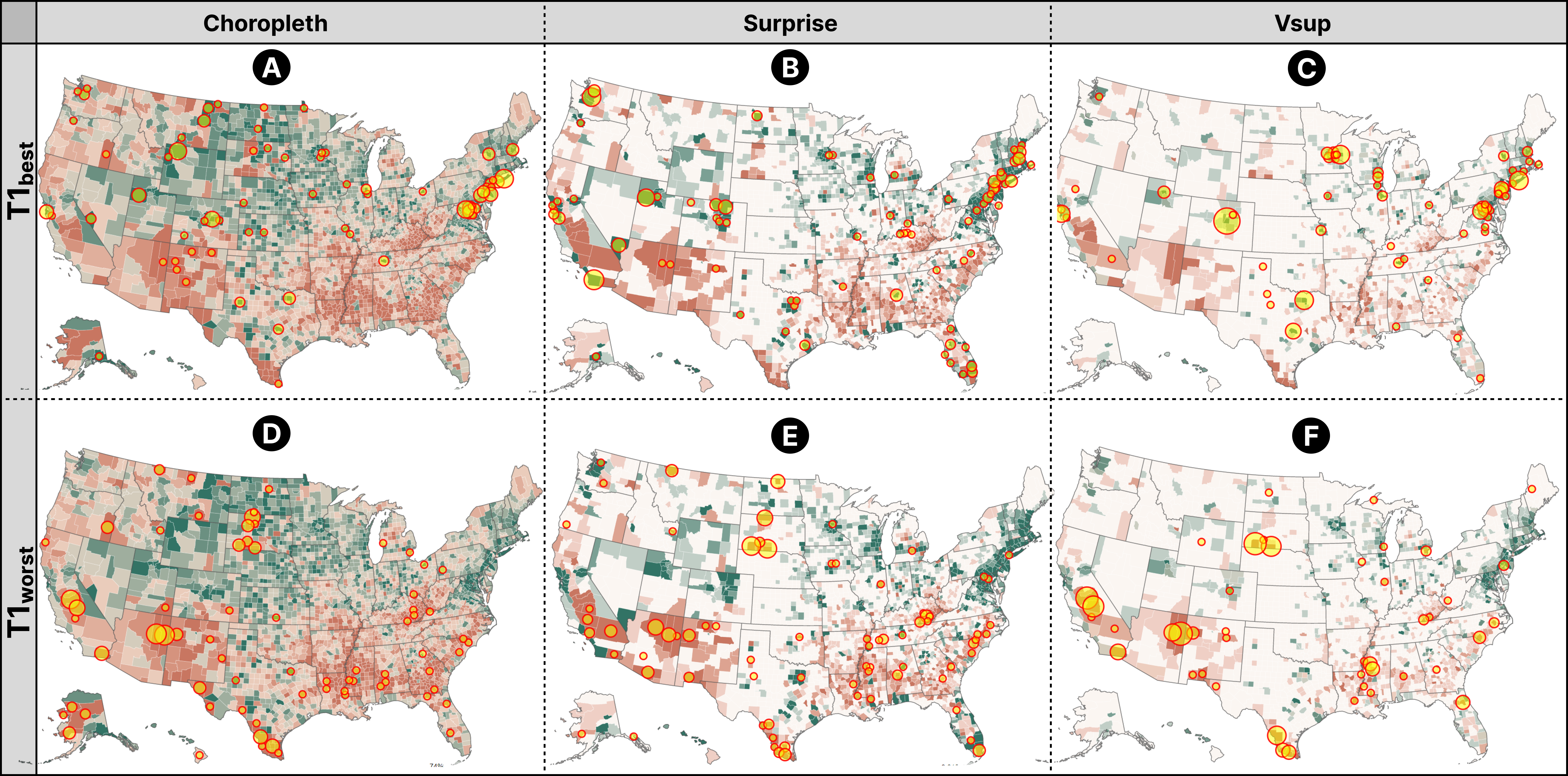}
    \caption{Participants' county selections for poverty data tasks T1\textsubscript{Best} and T1\textsubscript{Worst}. \textbf{A)} Choropleth map (T1\textsubscript{Best}) \textbf{B)} Surprise map (T1\textsubscript{Best}) \textbf{C)} VSUP (T1\textsubscript{Best}) \textbf{D)} Choropleth map (T1\textsubscript{Worst}) \textbf{E)} (T1\textsubscript{Worst}) Surprise map \textbf{F)} VSUP (T1\textsubscript{Worst}). Visual analysis shows a high degree of consensus on the VSUP maps compared to both the Choropleth and Surprise. The lack of consensus in Choropleth in this dataset compared to vaccine dataset may be due to skewed rate.}
    \label{fig:supplement-map}
\end{figure}

\newpage

\section{Sup-3: Quantitative results for rate, population and surprise metrics}
\begin{figure}[h!]
    \centering
    \includegraphics[width=0.95\textwidth]{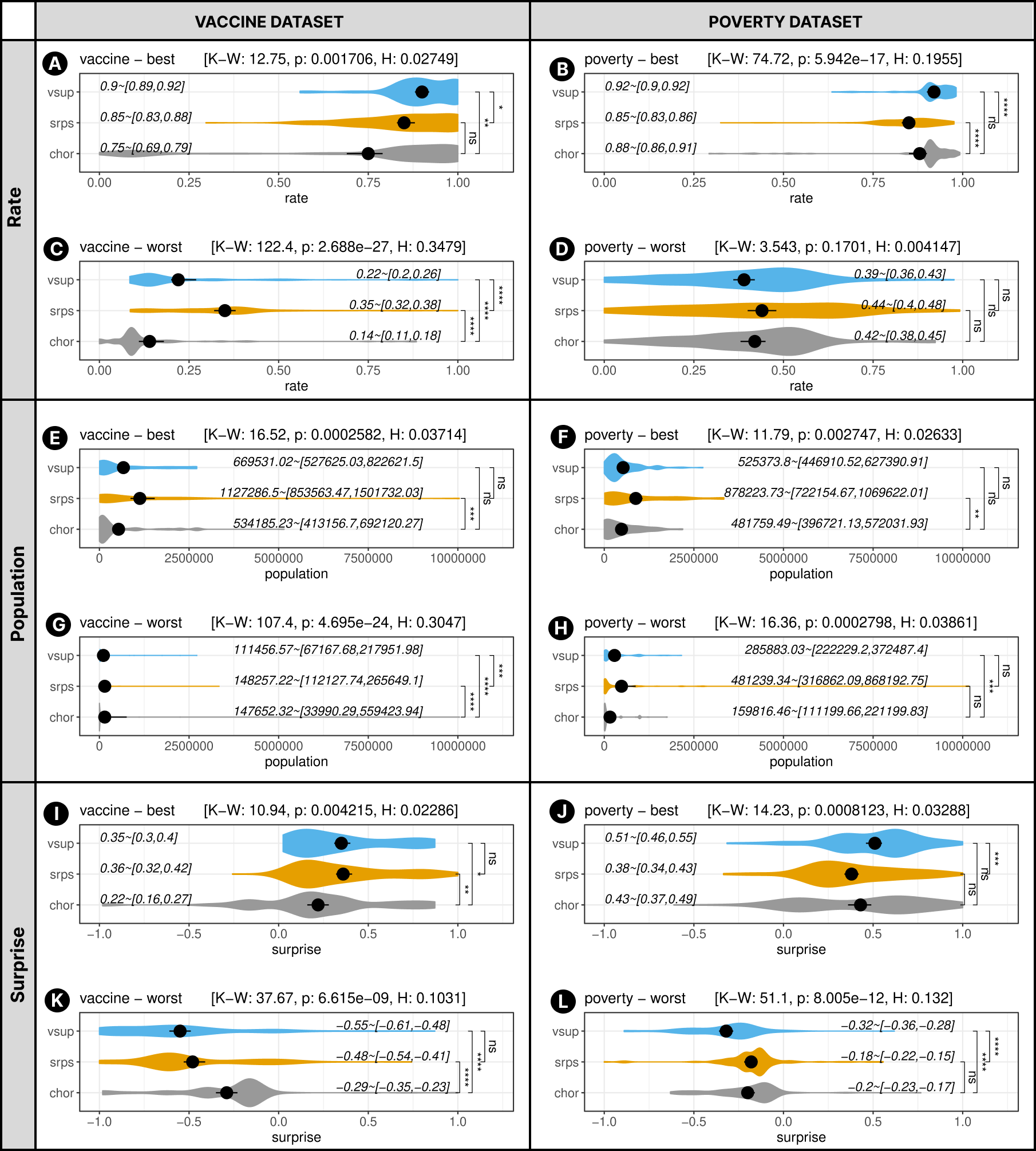}
    
    \caption{Quantitative results for rate, population and surprise metrics for both dataset and all conditions. Left column is based on vaccine dataset and right column is based on poverty dataset.
    \textbf{A)} T1\textsubscript{Best}-Vaccine (Rate) 
    \textbf{B)} T1\textsubscript{Best}-Poverty (Rate) 
    \textbf{C)} T1\textsubscript{Worst}-Vaccine (Rate) 
    \textbf{D)} T1\textsubscript{Worst}-Poverty (Rate) 
    \textbf{E)} T1\textsubscript{Best}-Vaccine (Population) 
    \textbf{F)} T1\textsubscript{Best}-Poverty (Population) 
    \textbf{G)} T1\textsubscript{Worst}-Vaccine (Population) 
    \textbf{H)} T1\textsubscript{Worst}-Poverty (Population) 
    \textbf{I)} T1\textsubscript{Best}-Vaccine (Surprise) 
    \textbf{J)} T1\textsubscript{Best}-Poverty (Surprise) 
    \textbf{K)} T1\textsubscript{Worst}-Vaccine (Surprise) 
    \textbf{L)} T1\textsubscript{Worst}-Poverty (Surprise) 
    We use the Kruskal-Wallis test to find differences in the significance of the data collected using the stimuli. We calculated a 95\% confidence interval using a bootstrap method. Quantitative analyses suggests that VSUPs lead to the selection of counties with high rates and high surprise whilst Surprise maps lead to the selection of highly populated counties.}
    \label{fig:supplement-stats}
\end{figure}

\newpage

\section{Sup-4: Stimuli design considerations}

\begin{table}[h]
    \caption{List of the design consideration for the experiment stimuli}
    \begin{tabularx}{\textwidth}{r X l}
        {\bfseries SN}&{\bfseries Design Consideration}&{\bfseries Design Element}\\
        \hline
        1 & When a participant hovers over a county, we display a tooltip showing an event rate, a surprise value and county population. & tooltip\\
        2 & Hovering over the legend, highlights all counties with a similar color encoding. & legend \\
        3 & To improve searching capability \cite{roth2013empirically}, we enable map zooming (x2) and panning (up, left, down, right). & map\\
        4 & We use a discrete scale from the D3 library (d3.scaleQuantize), to map domain values to a corresponding color. & color scale\\
        5 & We use the Interquartile range (IQR) of each dataset to specify the minimum and maximum values of each scale domain. & color scale\\
        6 & We set the size of each map to $950 \times 525$ pixels. & map \\
        7 & We use the geoAlbersUSA map projection. & projection\\
        8 & To mitigate visual distortion or misinterpretation of the maps, participants were asked to use either a laptop or desktop. & device\\
    \end{tabularx} 
    \label{tab:taskList}
\end{table}

\clearpage

\onecolumn{

}
\bibliographystyle{abbrv-doi}
\bibliography{manuscript}